\documentclass[aps,prd,preprint,groupedaddress]{revtex4-2}
\usepackage{latexsym,amsfonts}
\usepackage{hyperref}
\usepackage{graphicx}
\begin{document}    

\title{A geometrical theory of thermal phenomena based on the kernel of the evolution equation}

\author{Yuri V. Gusev}

\affiliation{Department of Physics, Simon Fraser University, 8888 University Drive, Burnaby, B.C. V5A 1S6, Canada}

\date{\today}

\begin{abstract}
The kernel of the evolution equation is used to build a mathematical theory of thermal phenomena of gaseous and condensed matter.  The group velocity of sound and the molar density are proposed to be its two thermal variables that replace the over-complete set of temperature, pressure and volume. The defining  constants of the New SI (2019) of physical units are used to form a dimensionless variable of the new geometrical thermal theory. The dimensionless thermal sum generates, by the variation principle, equations on physical observables. Energy and entropy are  neither defined nor needed in the present formalism. The molar specific heat is derived and its temperature dependence is verified with the experimental data for mono-atomic gases. Critique of the postulates and notions of traditional statistical thermodynamics is given.

\end{abstract}

\maketitle
 
\begin{flushright}

{\em ''One of the principal objects of research in my department of knowledge is to find the point of view from which the subject appears in the greatest simplicity.''} J. Willard Gibbs.

{\em ''The search for truth should he the goal of our activities; it is the sole end worthy of them. \ldots To find the one \ldots it is necessary to free the soul completely from prejudice and from passion; it is necessary to attain absolute sincerity.''} Henri Poincaré \cite{Poincare-book1913}, p. 201.

\end{flushright}

The finite temperature field theory has been recently derived using the kernel of the evolution equation \cite{Gusev-FTQFT-RJMP2015}. This mathematical formalism was applied to building a phenomenological theory of the specific heat of condensed matter  \cite{Gusev-FTSH-RJMP2016,Gusev-RSOS2019,Gusev-SHC-EV-arxiv2020}. The initial expectation was that this method could only work for condensed matter systems, whose physical properties differ drastically from those of gases. However, a further comparison of the thermodynamics of gases the new method showed that it could be also applied to gaseous matter. Therefore, the finite temperature field theory is a seed that can grow up to become the comprehensive theory of thermal phenomena, thereby replacing the traditional statistical thermodynamics, which is obsolete and inconsistent. The proposed theory can resolve old controversies within thermodynamics by replacing  it with the first fully {\em geometrical} theory.

The key points of the proposed geometrical theory of heat are: 
\begin{enumerate}
\item
its main functional, {\em the universal (thermal) sum}, is dimensionless and derived from the non-trivial topology of the Euclidean four-dimensional spacetime;
\item
the universal (thermal) sum depends on dimensionless variables that are determined by observed physical variables, {\em the group velocity of sound} and {\em the molar density};
\item
the universal sum generates, by {\em the variational principle}, equations for measurable physical quantities as functions of the observables, e.g. the specific heat;
\item 
the notions of energy and entropy are neither defined, nor needed.
\end{enumerate}

To provide a motivation for the readers, we first give the critique of statistical thermodynamics, however, this critique was not the reason for the new theory's conception.

\section{The inconsistency of statistical thermodynamics} \label{crtitique}

Apart from classical mechanics, thermodynamics  is the oldest physical theory. The active development of this subject occurred in the 19th century and lasted to the 20th century. Its history is extensively described and studied, e.g. \cite{Truesdell-book1980,Fenn-book1982,Muller-book2007}, or a brief overview  \cite{Bordonia-EPJH2013}. The number of thermodynamics textbooks is countless, see for example some classic \cite{Clausius-book1879,Planck-book2010,Planck-book1903,Gibbs-book1902,Guggenheim-book1933,Kennard-book1938,Poincare-book1908} and modern texts \cite{Schredinger-book1948,Landau-V5,Kubo-book1968,Kittel-book1969,Gaskell-book2018}. Thermodynamics used to be considered the most universal theory, applicable to all branches of physics, including cosmology, e.g. \cite{Tolman-book1934,Weinberg-book2008} and astrophysics, e.g. \cite{Zeldovich-book1971}. However, this universality is greatly exaggerated. In reality thermodynamics is not compatible with the rest of theoretical physics. 

Thermodynamics  is also one of the most trusted physical theories, and many prominent physicists of the past believed it would never change. However, every physical theory eventually reaches its limits of applicability and longevity, and thermodynamics is not an exception. In fact, attempts to modify or amend thermodynamics have never ceased, e.g. \cite{Truesdell-book1984,Muller-book1998,Andresen-book1983,Andresen-ACIE2011,Kozlov-book2002}. Every such approach had its own modification to implement. 

In contrast, we believe that instead of making modifications or amendments of thermodynamics, the existing theory of thermal phenomena should be entirely re-derived. A new theory should  be based on different physical principles, employ different mathematics and operate with different physical variables. This proposal  makes the theory of thermal phenomena a  fully geometrical theory, thereby reaching the long-existent goal of mathematical physics. It also replaces the present {\em thermostatics} with a dynamical theory that possesses physical time via its variables and its geometrical setting. Thus, it can describes physical {\em processes} instead of dealing with non-existent {\em states}. We suggest that in order to make a thermal theory {\em dynamical}, i.e. dependent on time, one should exclude the explicit use of both, temperature and time, as improper variables.

Statistical thermodynamics inherited several shortcomings from its very inception. These intrinsic problems persisted to the present time because thermodynamics was  viewed a universal theory  applicable to all fields of physical phenomena, while in reality the scope of applications of statistical thermodynamics is limited to the  properties of ideal (dilute monoatomic) gases \cite{Truesdell-book1980}. Some of these inconsistencies were noticed and discussed long time ago. This led to the criticism of statistical thermodynamics, e.g. work \cite{Landsberg-PB1964} concluded that 'conventional thermodynamics is not a proper deductive system' and suggested that the {\em geometrization} of thermodynamics could lead to a proper theory. Yet, the concept of geometrization, understood as the adoption of mathematical methods of modern geometry, is not sufficient, thermodynamics as a branch of experimental physics must be reformulated as well.

Let us list the inconsistencies of traditional statistical thermodynamics.
\begin{enumerate}
\item
The use of {\em two different kinds of thermodynamic functionals}, 
energy-wise thermodynamic potentials, defined axiomatically as functions of thermal variables, and dimensionless entropy defined by statistical methods.
\item
The  historical {\em ad hoc choice of its thermal variables}, which coincide with the redundant set of its observables - temperature, volume and pressure.
\item
The physically non-existent concepts of {\em equilibrium states} and {\em adiabatic processes} that lack physical time, which is still present in the theory implicitly. 
\item
The main equation of thermodynamics - {\em the equation of state - is only a constraint} on its variables, because it lacks any derivatives. 
\item
Thermal systems' {\em boundaries are not included explicitly}, even though they are present -  implicitly in the theory and physically in the experimental setup.
\end{enumerate}
Let us elaborate on these items.

\subsection{Redundant thermodynamic variables, temperature} \label{variables}

The first majour trouble of thermodynamics is its ad hoc physical variables. They are based on the historical development of thermodynamics, as a thermal theory of ideal gases. The triad of temperature, $T$, pressure, $P$,  and volume, $V$, was adopted during the epoch of invention of {\em steam engines}.  At that time, principles for building theoretical models of physical phenomena and modern mathematical methods were not discovered yet, therefore, thermal variables were chosen arbitrarily. The three {\em observables} listed above were selected for the study of a special kind of matter under special physical conditions, i.e. an ideal gas under conditions relatively close to ordinary environment. Later they were adopted and used for the description of other physical systems that are composed of very different media (condensed matter), under very different conditions, sometimes called 'extreme', e.g. \cite{Fortov-book2011}. Furthermore, these physical  observables became  {\em variables} of the thermal theory, even though there is no {\em a priori} reason they should be, and the operating roles of observables and variables were mixed up.

Volume and pressure are determined by mechanics and  measured by mechanical instruments.  Furthermore, volume is a geometrical parameter of the exterior (a vessel) of a thermal system; it is not an intrinsic characteristic of a physical system's medium (gas). Pressure is a characteristic of mechanical effects exerted by gas on the system's boundaries (the walls of  a vessel) and vice verse, it is not an intrinsic property of a medium. 

Only temperature is a new physical quantity introduced in thermal physics. Whether it is an intrinsic (fundamental) characteristics in comparison with the notion of heat was a long controversy, e.g. see van der Waals' decision between 'caloric' vs 'thermic' foundations for thermal physics. To the present day, temperature remains an empirical physical quantity - a derivative one according to the Revised SI (2019) \cite{Stock-Metro2019}, whose full range of accessible values is covered by stitched patches of various thermometry techniques that rely on different physical phenomena.

One of the first and still commonly used   tools for measuring temperature is a liquid-in-glass thermometer \cite{Fischer-RPP2005}. It employs a {\em mechanical} way of determining $T$, which is based on thee properties of thermal expansion  of liquid (ethanol, mercury, toluene etc) placed in a glass bulb. The linearity of this thermal expansion is {\em postulated} to define the linear scale of temperature. The difficulty of designing, making and calibrating such devices was obvious for centuries \cite{Gauvin-AS2012}, thus, until the most recent decades, scientists realized that temperature has no absolute meaning and definition. Finally, this understanding has been implemented in the Revised SI (2019), which introduced the fixed defining constant of the unit of temperature, known as the Boltzmann constant \cite{White-Metro2015}.

However, temperature is {\em not continuously} determined in the whole  range of  accessible temperatures. The International Temperature Scale ITS-90 above about 1 K \cite{Preston-Thomas-Metro1990} is covered with four physically {\em different} thermometry techniques \cite{BIPM-ITS-90}: radiation thermometers above 1235 K, platinum resistance thermometers between 14 K and 1235 K, gas thermometers between 3 K and 25 K, and vapour pressure technique at the liquid helium temperatures between 0.65 K and 3.2 K. The ITS-90 temperature scale contains as many as 17 {\em fixed points} defined by various phase transition phenomena \cite{Fischer-RPP2005,Bedford-Metro1996}. The determination of temperature is especially difficult below 1 K \cite{Schuster-RPP1994}, where measurements by several methods aere not reconciled with each other \cite{Fellmuth-Metro1993}, thus, only the Provisional Low Temperature Scale of 2000 (PLTS-2000) was accepted \cite{BIPM-PLTS-2000}. 

Summarizing, the temperature scale is empirically defined and the physical nature of temperature is ambiguous. This fact is in agreement with the Revised SI (2019) of physical units, where the unit of temperature, {\em kelvin}, depends on the definitions of three other units and is determined with the above mentioned {\em mise en pratique} methods \cite{BIPM-SI2019-kelvin}. 

The act of measurement defines a physical quantity, not the act of assigning an abstract concept to a mathematical symbol. It was a matter of accepted convention to assign the  scale of temperature to the linear scale of thermal expansion of liquids or resistance of metals.  This means that one actually measures temperature in the units of length (metre) in the former case and in the units of resistance (ohm) in the latter case, multiplied by appropriate physical constants. Let us repeat, the linearity of this scale is only postulated.

Some authors rejected temperature as a distinct physical quantity and replaced it with energy \cite{Landau-V5}, arguing it was a more fundamental quantity. However, this substitution contradicts the foundation of modern physics laid down by metrology \cite{Stock-Metro2019}, where the unit of temperature is defined by the Boltzmann constant \cite{BIPM-NewSI}. The identification of temperature with energy could only be done within physical kinetics of J.C. Maxwell \cite{Maxwell-book2011}. This equivalence is a direct consequence of the molecular theory, whose constituents are {\em massive particles} that obey the laws of classical mechanics \cite{Poincare-book1908}, Chap. . Therefore, it is not universal and it does not hold for other kinds of energy. Indeed, several temperatures for different kinds of energies are commonly defined and used in condensed matter and plasma physics. Besides, the notion of energy itself is troublesome as discussed below.

The fatal handicap of classical and quantum theories of matter  is the use of the absolute zero of thermodynamic temperature, $T=0$. Even though the absolute zero temperature cannot be attained experimentally and it does not exist theoretically in statistical thermodynamics, it is  used in the electronic theory of condensed matter. The quantum field theory for condensed matter physics is first formulated at zero temperature, and then its version at finite temperature is re-derived. The introduction of temperature changes  the number of physical variables in a theory. Therefore, this action substitutes one theory with another, different,  theory, which is not even related to the original one, because its limit at $T=0$ does not exist. 

\subsection{Critique of the equilibrium state} \label{equilibrium}

Another trouble of thermodynamics is the concept of the {\em equilibrium state} of a thermal system. This notion exhibits two intrinsic inconsistencies of thermodynamics: 1) it assumes that a system is independent of its exterior (isolated), 2) it describes a static system, independent of physical time. However, any thermal system co-exists with its surrounding because this is the only way to observe it and to implement changes of its state.  In thermodynamics, a thermal system is allowed to change its state, but this action contradicts its very definition: any physical {\em process} is time-dependent, thus, time must be, explicitly or implicitly, present in theory's equations, and the rate of change ('speed') of a system can only be determined with respect to the process of another system. The proposed pseudo-relativistic theory in a four-dimensional {\bf space-time} resolves this problem.

Furthermore, a thermal equilibrium state is a theoretical abstraction, which never occur in Nature. The constant temperature of a thermal system in equilibrium is either an approximation (a system cools down or heats up at a rate, which is slow in comparison to measurement), or it is  maintained at constant temperature by heating (cooling), which compensates the energy loss (gain). Thus, the thermal equilibrium state is not really static, it is a special case of the dynamical process, when an outgoing energy flux is (almost) compensated with an incoming energy flux. This is it, no thermal system is ever closed and independent of time. This deficiency became apparent when recently thermodynamics  encountered  confrontations with experiment at very 'small' length scales (approaching the average inter-atomic distances) and/or very 'short' time periods (approaching the minimal time periods of acoustic waves supported by media). For an older example, one can look at the classic works on the Chambadal-Novikov efficiency of a heat engine \cite{Chambadal-book1957,Novikov-JNE1958}, where 'timeless' thermodynamics comes into conflict with experiment in regards to the efficiency of heat engines. It was rediscovered in Ref. \cite{Curzon-AJP1975} and popularized and elaborated in numerous works later.

\subsection{Critique of energy}  \label{energy}

For nearly two centuries the theory of thermal phenomena rested on the paradigm of two inseparable concepts of {\em energy and  entropy} \cite{Fenn-book1982,Muller-book2007}. These concepts form the double foundation of  {\em statistical thermodynamics}, even though they are entirely different in their mathematical constructions and physical meanings. Maxwell's statistical physics is a mathematical description of atomic systems, which is built on the mechanics of a thermal system's constituents (molecules) \cite{Maxwell-book2011}. In contrast, Gibbs' thermodynamics \cite{Gibbs-book1902} is a phenomenological theory built axiomatically on abstract notions of thermodynamic potentials and ensembles that are not directly related to experiment \cite{Kubo-book1968}. 

The functionals of thermodynamics, thermodynamic potentials, have the physical dimensionality of {\em energy} (measured in {\em joules} or $kg \cdot m^2 \cdot s^{-2}$): internal energy, free energy, enthalpy etc. They are macroscopic (global), i.e. assigned to the thermal system as a whole, phenomenological functionals.  In addition to them, there is one {\em dimensionless} functional, called {\em entropy}. Dimensionless entropy is made the energy-like multiplier  of the Boltzmann constant and temperature $k_{\sf B} T$. The introduction of entropy was needed in order to make thermodynamics, otherwise a physically incapable model, a theory capable of describing physical phenomena, understood mainly as the theory of heat engines \cite{Carnot-book1897}. 

Furthermore, as different from thermodynamic potentials, entropy is determined by the {\em statistics} of constituents of a matter system.These two kinds of thermal functionals come from two different branches of mathematics: the theory of continuous functions and the probability theory. This union of statistical physics and thermodynamics is traditionally called statistical thermodynamics \cite{Schredinger-book1948,Guggenheim-book1956} (see the long title of the last monograph by J.W. Gibbs \cite{Gibbs-book1902} that served as a theoretical foundation for these books). Let us note that it took thirty years \cite{Guggenheim-book1933} and the nearly thousand pages commentary \cite{GibbsCommentary-book1936} for Gibbs' ideas to become accepted.

These two concepts so deeply penetrate modern physics that it seems impossible to construct physical theories or even think about physics without them. However, the physical notion of energy is really ambiguous as was emphasized by many, for example, by E. Schr\"odinger \cite{Schredinger-INC1958}. Only one further step towards discarding the quantity of energy was left - finding  a way to describe thermal phenomena {\em without} the notion of energy. It can only done after realizing that energy, like entropy, is {\em not} an observable physical quantity, it is an abstract concept of physical theories. With the progress of science, physicists gain new mathematical tools and gather more experimental facts, which allow them to stop using obsolete notions, after all, the theory of {\em phlogiston} was also in agreement with older observations. 

However, an absolute value of energy is not even defined in physical theories, including classical mechanics, only the difference of energy has a physical meaning. However, changes (differences) of energy-like physical quantities can occur, explicitly or implicitly, only over finite periods of time. Therefore, any experimental equipment really measures the rate of the change of energy, i.e. {\em power}, like the output of $I\cdot V$, electric power. The power is always determined and measured locally, i.e. per unit of an area of the system's boundary, in other words, the {\em energy flux} is a true physical observable. 

Furthermore, statistical thermodynamics operates with the notion of discrete (exact) values (often called levels or states) of energy. Of course, this is over-idealization, because no physical quantity can be measured without an experimental uncertainty. Ideally discrete levels of energy  (states) do not exist, as any energy state possesses the uncertainty ('line width') that could be determined by an experiment of higher precision. Historically, the widespread use of energy states was brought in by the convenience of a mathematical apparatus available for the evaluation of (infinite) sums. We argue that the only way to introduce truly discrete  quantities into physics is via topology. This is it, in geometry, the only discrete quantity is the number of windings of a geodesic line along the cyclic Euclidean time \cite{Gusev-FTQFT-RJMP2015}. Topologically, this corresponds to the number of holes of a manifold \cite{Fomenko-book1992}. We will use the fact that physical discreteness has topological nature in the proposed theory of heat.

\subsection{Critique of entropy}  \label{entropy}

The concept of entropy is one of the most ubiquitous and at the same time mysterious notions of theoretical physics \cite{Chambadal-book1963}. The historical reason for the introduction of entropy was a disagreement between results of molecular physics and axiomatic thermodynamics, e.g. \cite{Gaskell-book2018}. Namely, the thermal energy of an ideal gas was assumed to be equal to the total {\em mechanical} (kinetic) energy of its molecules. It is instructive to note that the pioneers of thermal theory, such as R. Clausius \cite{Clausius-book1879} wrote about a 'mechanical' theory of heat, as this intrinsic concept really confine the scope of applicability of thermostatics and statistical physics. Therefore, it was expected that the change of this kinetic energy should be equal to the mechanical work done by a system plus the dissipated heat. This equality was not experimentally observed. To express this fact quantitatively the notion of entropy was introduced, and the observed {\em discrepancy} was expressed as {\em the second law of thermodynamics}. The very origin of the name 'free energy' is to mean the heat energy of a thermal system that is available ('free') to use for practical (mechanical) purposes. Free energy is really introduced into thermodynamics with help of entropy. However, this is a typical way of making amendments to a theory, which is intrinsically inconsistent, by introducing additional quantities: more free parameters can help to fit any theory to experiment. 

It is true that getting rid of false scientific dogmas is hard. Let us quote E. Schr\"odinger again \cite{Schredinger-INC1958}:{\em ''Yet I do not think that it has ever occurred to anyone to declare that entropy is {\rm not} a property of a physical system {\rm per se}, but only expresses 'my' knowledge about that system''}. Yet, entropy is {\em not} a measurable quantity (property) of physical systems, therefore, it may or may be not used in physical theories. We argue that theories with {\em no entropy} are simpler, have a wider range of applications and a more predictive power than textbook statistical thermodynamics, whose time to retire is long over due.

A statistical expression for the entropy was introduced in physics using combinatorics, Boltzmann entropy was later matched by Shannon entropy, which fuelled the activity in the information theory. However, it became obvious that the statistical definition of entropy is not unique. A large number of definitions of entropy  exist within theoretical physics, so-called 'entropy zoo' \cite{Faist-arxiv2016}, pp. 64-65. This is the glaring proof of a faulty status of entropy in physics. 

Theoretical physics should generalize and simplify the description of physical phenomena that can be measured or defined by physical observables. If a new concept makes physics more complicated and does not increase its predictive power, then it should not be used. In the proposal below there is no need for introducing entropy, i.e. entropy is not needed to describe physical phenomena by the finite temperature field theory. There are simply nor mathematical methods neither physical reasons for the existence of entropy.

In the textbooks on statistical physics, e.g. \cite{Landau-V5}, the derivative of free energy over temperature is declared entropy. This definition was copied by the finite temperature quantum field theory \cite{Kapusta-book2006}. However, the adoption of this definition is questionable because the latter theory deals with the creation and annihilation of  particles, while entropy is determined by the means of statistical physics developed for systems with a constant number of particles. The introduction of chemical potentials, similar to chemical potentials in thermodynamics, into finite temperature quantum field theory to deal with a variable number of particles amounts to adding mass-like parameters to this theory. Thus, while chemical potentials remove intrinsic, non-physical, divergences of QFT, they also eliminate its predictive power, which is actually null.  

Using the notions of the finite temperature field theory \cite{Gusev-FTQFT-RJMP2015}, which were inspired by but not derived from the finite temperature QFT, we suggested that the functionals of thermodynamics are dimensionless, like entropy, but unlike entropy they are not defined by statistical methods. We suggest that the evolution of a thermal physical system is determined by a functional we called the universal thermal function. Historically, the name 'free energy' was used \cite{Gusev-FTQFT-RJMP2015,Gusev-FTSH-RJMP2016}, however, this functional is completely different from the Helmholtz and Gibbs free energy in the meaning and in the definition. Therefore, the new name should reduce the confusion by emphasizing that we develop a new theory that is not  derived from or related to ordinary statistical thermodynamics.

The notion of entropy is intrinsically connected with the second law of thermodynamics. However, neither entropy nor the second law is needed in the proposed new framework. Needless to say, numerous problems and even mystiques related to the second law of thermodynamics used to be known to a long time, no wonder the book \cite{Capek-book2005} ends up with the conclusion: {\em ''the second law is} famous for being famous. {\em It has become the epitome of scientific truth, notorious for its absolute status, virtually unquestioned
by the scientific community''}.

\subsection{Critique of the equation of state}  \label{EOS}

The three observables of thermodynamics are connected together by {\em the equation of state} (EOS) for ideal gases. Its long history is reviewed in the cited books, here we reproduce it as,
\begin{equation}
	pV = n R T,  \label{pvt-R}
\end{equation}
where $n$ is the number of moles, and $R$ is the universal gas constant. This {\em universal} equation, i.e. valid for many gases, was first introduced by Dmitri Ivanovich Mendeleev, who is famous for the periodic table of chemical elements, but spent many years researching experimentally and theoretically the physics of gases. Mendeleev reported it first on Sept. 12, 1874 \cite{Mendeleev-JRCPS1874} and published in English in {\em Nature} \cite{Mendeleev-Nature1877-2}. Mendeleev discovered this {\em universal} gas constant, in contrast to specific constants for various gases, by combining three laws of ideal gases - Boyle-Mariotte law, Gay Lussac law and Amper-Gerland law. Eq.~\ref{pvt-R} generalizes the Clapeyron equation (valid for specific gases with specific gas constants) and is sometimes called the Mendeleev-Clapeyron equation. Furthermore, he measured the universal gas constant with high precision, within $0.3\%$ of its modern value \cite{Mendeleev-Nature1877-2}. Following an earlier proposal \cite{Kalafati-HNST1984}, we call $R$ {\em Medeleev gas constant}, in an agreement with the scientific tradition for naming physical constants by the names of discoverers: Boltzmann constant, Avogadro constant and Planck constant. Thus, D.I. Mendeleev was the first scientist, who realized that physical laws of gases should be expressed in molar units (per mole), thus making a great step towards further studies by van der Waals and then modern theories of the scaling.

All deficiencies of thermodynamics are exposed by this equation, which is supposed to describe the {\em evolution} of a thermal system in terms of its thermal variables. In reality, this is a {\em static} equation, where all three thermal variables enter linearly (or polynomially). For all practical purposes, the empirical EOS discovered by van der Waals \cite{Waals-thesis1873,Sengers-book2002,Kipnis-book1996} remains the working tool of engineers and experimental physicists, e.g. \cite{Sengers-book2000},
\begin{equation}
	\left( p + \frac{n^2 a}{V^2}\right) ( V - n b ) = n R T,  \label{Waals}
\end{equation}
where $a$ and $b$ are dimensional calibration parameters.

It is easy to see that the EOS for ideal gases (\ref{pvt-R}) or real gases (\ref{Waals}) is really a {\em constraint} on physical parameters of a thermal system, which reduces the {\em redundant set} of three thermal variables to actual two. It is not a dynamical equation of physics, because it does not not contain any derivatives over physical variables and does not depend on physical time at all. For these two reasons, it cannot describe the dynamical behaviour of a thermal system and consequently cannot predict the system's evolution. In practice, the equations of state are empirically fitted from large sets of experimental data for physical parameters, e.g. \cite{Zharkov-book1968}. In theoretical physics the situation ought to be opposite: proper physical equations are supposed to describe and predict behaviour of physical quantities.

\section{Introduction to the new theory}

The new mathematical formalism  and its application to the study of thermal properties of real gases are outlined below. We argue here that a mathematical theory of thermal phenomena can be formulated  in terms of continuous functions only, without supplementing it with statistical concepts, if the theory is formulated in the four-dimensional space-time.

\subsection{Motivation for geometrical formalism}

This proposal is not related to previous attempts of the {\em formal geometrization} of thermal theory. Reshaping and reformulating the statistical thermodynamics cannot cure its ills, thus, these attempts are futile as proven by the long history of such efforts, e.g. e.g. \cite{Caratheodory-MA1909,Weinhold-JCP1975}. The present work is an outcome of the axiomatic study that emerged after clearing mathematical errors and physical contradictions of the finite temperature quantum field theory in the effective action method \cite{Gusev-FTQFT-RJMP2015}.

One of its main  postulates \cite{Gusev-FTQFT-RJMP2015} states that no physical theory of matter can be formulated at zero temperature. Indeed, the limit $T=0$ is not attainable either experimentally or theoretically. The commonly used 'low' temperature limit can  be physically understood and mathematically defined only as the asymptotic, $T_{\mathrm{ref}}/T \gg 0$, with some reference temperature, $T_{\mathrm{ref}}$, the concept introduced in the solid state physics by P. Debye \cite{Debye-AdP1912} and in the liquid-gas state physics by van der Waals \cite{Waals-thesis1873,Sengers-book2002,Kipnis-book1996}. However, any reference  temperature, e.g. the temperature of melting or the threshold of the quartic power law of specific heat, is material specific, i.e. different for different substances. Because the range of  'low' temperatures, on the absolute temperature scale, varies significantly for various  materials, the notion of 'low temperature' is ambiguous and not acceptable in physical theories. Therefore, one can only define and use the notion of {\em quasi-low temperature} (QLT) \cite{Gusev-RSOS2019,Gusev-SHC-EV-arxiv2020}, when the traditional scale of thermodynamic temperature is abandoned. 

To develop a thermal theory (nearly) universally applicable to various matter systems, and even to different phases of matter, one should make variables of its equations {\em dimensionless}.  This means its physical equations ought to become {\em mathematical} expressions, whose variables are {\em pure numbers}, with null physical dimensionality. Then, thermal phenomena of different  matter system can be studied with the same equations and {\em calibrated} (connected with physical measurement) by a single (material specific) parameter. 

Dimensionless variables should be determined by {\em observable} quantities, because they enter a {\em phenomenological} theory.  Since naive definitions, like the ratio, $T/T_{\mathrm{ref}}$, are not acceptable, dimensionless variables should be combinations of  the {\em defining}  physical constants and various physical characteristics of a matter system. Let us emphasize that we separate physical {\em observables} from physical {\em variables}. Their coincidence is an anachronism of simple theories of the past. The pair, the group velocity of sound in a medium and the average inter-atomic distance (equivalently, the molar density), incidentally discovered in \cite{Gusev-FTSH-RJMP2016} is a good (but perhaps not the only) candidate to serve as such observables.

Let us emphasize that only dimensionless solutions of dimensionless equations can be mathematically coherent, because only such expressions can produce limits and approximations, e.g. series expansions, that are physically and mathematically consistent. Indeed, the convergence and the limit of validity of any expansion in a {\em dimensionful} parameter is impossible to quantify. The ubiquitous case in physics is the  expansion in the powers of the Planck constant, which is usually justified by its small absolute value, $h = 6.626 070 15 \cdot 10^{-34}\ {\mathrm J}\cdot {\mathrm s}$ \cite{CODATA-Metro2018}. However, the coefficients of the terms in this expansion can be arbitrarily large, and in quantum field theories they are claimed to be infinite \cite{DeWitt-book2003}, which is physically meaningless. 

In theoretical physics there exists a myth that the presence of the Planck constant in an equation signifies some magic 'quantum' properties of its solutions. However, the Planck constant was originally introduced by M. Planck within classical thermodynamics \cite{Planck-book1914,Planck-book2010}, while at present time, within the New SI (2019) of physical units \cite{BIPM-NewSI}, which is based on the defining physical constants \cite{Stock-Metro2019}, the Planck constant {\em defines the unit of mass} \cite{Wood-AdP2019}. This is the only physical meaning of the Planck constant in physics as an experimental science.

Here we propose the geometrical implementation of a thermal theory based on the evolution equation, which is the fundamental equation of geometrical analysis. This proposal consists of three elements: 1) utilizing the kernel of the evolution equation, 2) deriving the universal thermal sum from the topology of the {\em four-dimensional Euclidean spacetime}: the three-dimensional manifold with a boundary and the closed time coordinate, 3) using the defining  physical constants, including the Planck constant, to render physical variables dimensionless.

The presence of a boundary of a physical system is required for physical consistency of the theory. Indeed, the boundary defines the very existence of a condensed matter system and contains a gaseous matter  system, yet most physical theories, including thermodynamics, lack even the notion of boundaries. In the finite temperature field theory, it is possible to directly study the physics of boundaries \cite{Gusev-SHC-EV-arxiv2020}.

In recent decades, experiments with physical systems, which are deemed to be 'small', e.g. the size of nanometers in absolute length units, showed that the boundary effects maybe dominating physics of these system. The explicit use of boundaries also introduces another (global) length scale of a physical system, e.g. for the thermal radiation phenomena \cite{Gusev-FTQFT-RJMP2015} where the system's size may be not ''small'. The system's size as a length scale was long known and utilized in the fluid dynamics theories. 

Even a system, which consists of locally interacting constituents (molecules), exhibit physical phenomena that are {\em nonlocal}. Obviously, in a theory based on the properties of sound (elastic) waves in matter, global (nonlocal) solutions of the four-dimensional Euclidean wave equation, are important. Due to the high speed of sound in any matter, sound waves provide the fast attainment of the {\em thermal equilibrium} of gaseous matter. Sound waves and their interference also serve as a fundamental mechanism for the emergence of {\em thermal fluctuations}, which in traditional thermodynamics are declared without explanations.

\subsection{Measuring and defining the thermodynamic temperature} \label{temperature}

Temperature is a key observable of thermal phenomena, it is measured by a variety of experimental techniques \cite{Moldover-NP2016,Quinn-book1990,Michalski-book2001,Fischer-RPP2005}. There used to be two views on traditional thermodynamics, 'caloric' and 'thermic', the former one assumes heat is a fundamental quantity, while the latter one assumes temperature is fundamental. In any case, traditional thermodynamics assumes that thermal energy (heat), $E$, is stored in gaseous or condensed matter, and temperature, $T$, is the measure of energy, $E$, per  'degree of freedom' \cite{Moldover-NP2016}, which means that the equipartition theorem is valid: $E = 1/2 k_{\mathsf{B}} T$. 

However, the notion of a degree of freedom belongs to the mechanical  (atomistic) model of matter, while the equipartition theorem is a physical postulate whose experimental validity is limited to an ideal gas (rarefied monatomic gas). This postulate certainly does not hold for other types of matter, and even for an ideal gas its validity is limited to certain conditions. Therefore, this limited correspondence between energy (heat) and temperature, i.e. between mechanics and thermodynamics,  is not universal (fundamental). Without this link, a consistent thermal theory, with a predictive power, cannot be developed. We propose to abandon both, heat and temperature, in favour of the dimensionless thermal function expressed with a dimensionless variable expressed through physical observables.

The common  instrument for measuring the temperature, the liquid-in-bulb thermometer, operates as a geometrical device, as the temperature is defined by the thermal volume expansion of a liquid. However, if we recall that a boundary is needed and always present to define a thermal system, temperature is really the measure of {\em energy flux} from physical medium (material body or gas) to a measuring device, thermometer, through its common boundary (or the boundary of a vessel confining gas). 

In recent decades, the most precise method for the determination of temperature has been the measurement of a velocity of sound in rarefied noble gases. Therefore, the gas acoustics \cite{Moldover-Metro2014} has  been selected by the international metrology committee, CODATA, for the final determination of the Boltzmann constant \cite{Podesta-Metro2013} in the New SI (2019)  of physical units \cite{Stock-Metro2019}. The method is based on the equation of the sound velocity in dilute gas which holds over the wide range of temperatures \cite{Moldover-Metro2014}. Technical details of the acoustic thermometry can be found in the review \cite{Gavioso-Metro2015}. With the Boltzmann constant's value fixed after May 2020, the acoustic thermometers serve to measure temperature and determine the unit of temperature with ever higher precision.

Thermodynamic temperature is {\em assigned} to be linearly correlated with  the thermal expansion of liquid matter, but in reality this is a historical convention which determined temperature, temperature does not exist as a fundamental field-like quantity, as commonly assumed in modern mathematical theories of thermodynamics. 

In general, temperature may or may not have a linear correlation with acoustic velocity of thermal expansion.

The quantity of temperature is empirical and it is {\em de facto} defined (determined) by the group velocity of sound in the rarefied monatomic gas

Therefore, it is logical to abandon the notion of temperature and replace with the physical quantity which really defines it, the velocity of sound in matter. 

Let us look at the equation for the sound velocity \cite{Moldover-Metro2014},
\begin{equation}
v_s = \left(\frac{\gamma_0 R T}{M} \right)^{1/2},
\label{soundv}
\end{equation}
where $\gamma_0$ is the ratio of the gas specific  heats $\gamma_0 = C_p/C_v$  in the limit of low ('zero') pressure (for monoatomic gases, $\gamma_0=5/3$), $T$ is the thermodynamic temperature, and $M$ is the molar mass of gas. If we rearrange terms of the equation (\ref{soundv}) as,
\begin{equation}
M v_s^2 = \gamma_0 R T,
\end{equation}
it looks like the kinetic energy of a body with mass $M$ that moves with the speed of sound $v_s$. Then, it can be considered as the definition of temperature via the velocity of sound, as it is indeed used in the acoustic thermometry. This relation indicates the fundamental role of the velocity of sound in thermal theory as well as a special, mechanical, nature of this definition of temperature, which is limited to ideal gases under special conditions.

The lack of a variable connected with time was the crucial handicap of thermodynamics, which remained thermostatics. The velocity of sound in gaseous or condensed matter is the only intrinsic property of a system, which is coupled with physical time. Thereby, the velocity of sound establishes the global time scale of a thermal system, because it limits the speed with which changes in the system's molar density, called {\em fluctuations}, can propagate across a system. Because the velocity of sound is typically such that the global time scale is much shorter than the time scale of external changes, the physical time could have been neglected in equilibrium thermodynamics (but not in the heat conductivity). However, this is no longer the case for recent experiments  due to either small system sizes and high frequencies of external influence.

\subsection{The evolution equation} \label{evolution}

Theoretical physics must be {\em relativistic} \cite{Poincare-RCMP1906}. Therefore, any physical theory must be built in the four-dimensional {\em space-time}, introduced by H. Minkowski \cite{Minkowski-PZ1908}. The case of  electrodynamics was obvious and elaborated immediately by Lorentz, Poincar\'{e} and Einstein, but the application of this physical concept to thermal phenomena was delayed until the 21st century. 

The Planck constant was originally introduced by M. Planck in the context of thermodynamics, \cite{Planck-AdP1900,Planck-book1972,Planck-book1914} in order to find the empirical law of thermal radiation later named after him. Without one of four {\em defining} (formerly called 'fundamental') physical constants \cite{CODATA-Metro2018} it is impossible to construct  physical theories. In the New SI (2019) of physical units, the Planck constant is fixed as the defining constant of the unit of mass \cite{Stock-Metro2019}. We show how it can be used to connect mechanical (elastic) properties of matter with thermal ones \cite{Gusev-FTSH-RJMP2016,Gusev-RSOS2019}.

\subsection{Universal thermal function} \label{functionals}

\subsection{Dimensionless variables} \label{dimensionless}

In order to connect mathematics with experimental physics we need the defining physical constants. For reference, the molar gas constant, $R$, used to be defined by the Boltzmann's constant \cite{CODATA-Metro2018},
\begin{equation} 
k_{\mathsf{B}} = 1.380\, 649 \times 10^{-23}\ J K^{-1},
\end{equation}
and the Avogadro constant \cite{CODATA-Metro2018},
\begin{equation}
N_{\mathsf{A}}= 6.022\, 140\, 76 \times 10^{23}\ {mol}^{-1}, 
\end{equation}
\begin{equation}
R = N_{\mathsf{A}} k_{\mathsf{B}} = 8.314\, 734\ J\, mol^{-1}\, K^{-1}.
\end{equation}
In the New  SI (2019), the Boltzmann constant is determined by the measurement of the molar gas constant \cite{CODATA-Metro2018}.  

Planck's inverse temperature, whose physical dimensionality is length, is defined in a way that resembles its definition in finite-temperature field theory,
\begin{equation}
\beta = \frac{1}{B}\frac{h v_{\mathrm{s}}}{k_B T}, \label{beta}
\end{equation}
where $v_{\mathrm{s}}$ is the group velocity of sound and $B$ is the calibration parameter determined by experiment for the whole class of thermal systems, e.g. the 'ideal gas'. This equation should be really understood as a definition of thermodynamic temperature, $T$, because $\beta$ is an intrinsic mathematical variable of this geometrical theory, it is the length of the circumference $S^1$.

The dimensionless thermal variable for the geometrical theory of any matter systems is defined as,
\begin{equation}
\alpha \equiv \frac{\beta}{a} = \frac{1}{B}  \frac{\hbar v_{\mathrm{s}}}{a k_{\mathsf{B}} T}, \label{alpha}
\end{equation}
where $a$ is an average inter-atomic distance of a matter system, which can be replaced by the lattice constant for crystalline matter or determined via the molar volume, $V_{\mathrm{M}}$,
\begin{equation}
a \equiv \big(V_{\mathrm{M}}/N_{\mathsf{A}}\big)^{1/3}, 
\label{molarV}
\end{equation}
for gaseous systems. The parameter $a$ is the  characteristic length scale of a system, which naturally defines validity of the used physical approximation by limiting the minimal allowed length of sound waves in matter. The minimal wavelength plays a dual role of an intrinsic regulator of the theory's equations and a parameter needed to render the theory's variables dimensionless.

Let us now use the equation for the sound velocity (\ref{soundv}) to express thermodynamic temperature  as,
\begin{equation}
T = \frac{v_s^2 M}{\gamma_0 N_{\mathsf{A}} k_{\mathsf{B}} }.
\label{Tviav}
\end{equation}
Substituting Eq. (\ref{Tviav}) into (\ref{alpha}) we obtain a new definition for the dimensionless variable that does not contain temperature, the physical observable (and variable at the same time) we want to get rid of,
\begin{equation}
\alpha  = \frac{1}{B} \frac{\hbar N_{\mathsf{A}}}{M}  \frac{1}{a v_{\mathrm{s}}}. 
\label{alphaMv}
\end{equation}
Note, that this equation does not contain the Boltzmann constant as well, which should be really expected, as in the New SI (2019) $k_{\mathsf{B}}$ defines the unit of temperature. Eq. (\ref{alphaMv}) is factorized according to the physical meaning of its terms, only the last term changes along with the system's thermal evolution.
However, the average inter-atomic distance is not a physical observable, therefore, we use (\ref{molarV}) to replace it with the molar volume,
 \begin{equation}
\alpha  = \frac{1}{B} \frac{\hbar N_{\mathsf{A}}}{M}  \frac{1}{a v_{\mathrm{s}}}.
\end{equation}

\section{Summary}  \label{summary}

Let us summarize the studied topics.
\begin{itemize}
\item
The redundant set of three ad hoc thermal variables of statistical thermodynamics is replaced with the group velocity of sound and the molar density.
\item
In order to make the thermal theory dynamical, i.e. dependent on time, we should exclude the explicit use of both temperature and time, as improper thermal variables.
\item
The defining physical constants, including the Planck constant which defines the unit of mass, are used to form a dimensionless thermal variable.
\item
The unified mathematical theory of thermal phenomena for gaseous and condensed matter is based  on the kernel of the evolution equation.
\item
The unique thermal sum, a dimensionless functional derived in the topologically non-trivial four-dimensional spacetime, replaces the redundant and eclectic set of thermodynamic potentials, energy, entropy etc.
\item
The universal thermal sum generates the dynamical equations of thermal physics that replace the static constraint of statistical thermodynamics.
\item
Physical time is implicitly introduced into the theory, via the velocity of sound in matter.
\item
Neither entropy, nor energy is not needed for the geometrical theory of thermal phenomena.
\item
Boundaries of a physical system  are explicitly included in the theory due to its geometrical nature.
\end{itemize}

\section{Discussion}  \label{discussion}

{\em Conclusion of a theoretical analysis}
The motivation for the developing a new geometrical theory of thermal phenomena was the clarification and error correction in the finite temperature quantum field theory. However, the motivation for the writing of this paper could be better explained with the words of J.W. Gibbs \cite{Gibbs-letter1892}: {\em '' I am trying to get ready for publication something on thermodynamics from the a priori point of view, \ldots I do not know that I shall have anything particularly new in substance, but shall be contented if I can so choose my standpoint (as seems to me possible) as to get a simpler view of the subject.''} 

Indeed, the main goal of theoretical physics is the simplification of mathematical description of natural phenomena, based on the generalization of experimental facts. This idea is the foundation for all successful theories of physics. Contrary, the over-complication of a physical theory necessarily leads to the destruction of its predictive power. We hope that sketched the way to further simplification of the theory of thermal phenomena by reducing the number of its physical parameter from three to two and the number of its functions from several to one, while keeping its descriptive power. As for predictions of the new theory, they would emerge in the field of critical phenomena, which is a subject of another publication.

Logical inconsistencies of statistical thermodynamics have been known for a long time, however, the development of modern mathematical  theory of thermal phenomena was delayed by the lack of appropriate mathematical methods and the reluctance to discard the notions of energy and entropy.

The thermal field theory resembles the Gibbs theory of  thermodynamic ensembles \cite{Gibbs-book1902,GibbsCommentary-book1936} by the mathematical form of its equations. We believe this is not a coincidence, but modern mathematics explains the axiomatic definitions of J. Willard Gibbs, who was one of the greatest pioneers of modern physics \cite{Kraus-Science1939}. In contrast to statistical thermodynamics, in the proposed geometrical theory of thermal phenomena, there are no statistical notions emerging from the atomistic description of matter. The only discrete mathematical construction is the universal thermal sum. The theory's functional is dimensionless  $F(\alpha)$ with the dimensionless variable $\alpha$. For historical reasons, one has to interpret obtained mathematical results in terms of the traditional physical observable, temperature, this interpretation can be done via the definition of $\alpha (T)$. 

{\em The physical world is a four-dimensional space-time.}

The geometrical nature of thermodynamics was elucidated from its inception, when geometrical methods were introduced by Josiah Willard Gibbs  in his first thermodynamic papers \cite{Gibbs-TCA1873-1,Gibbs-TCA1873-1}. This connection was further elaborated in the works of James Clerk Maxwell \cite{Maxwell-book1902}. 

Shortly after statistical thermodynamics was completed by Gibbs \cite{Gibbs-book1902}, the four-dimensional nature of physical spacetime was discovered by Hermann Minkowski \cite{Minkowski-JDM1909}:
{\em ''The fact that the world-postulate holds without exception is, I would like to believe, the true core of an electromagnetic picture of the world; \ldots In course of further developing the mathematical consequences, enough suggestions will be forthcoming for the experimental verification of the postulate; in this way even those, who find it unsympathetic or even painful to give up the old, time-honoured concepts, will be reconciled by idea of a pre-established harmony between pure mathematics and physics.''} 

Obviously, electromagnetic phenomena underlie thermal ones: thermal radiation is electromagnetic radiation of certain frequencies and collisions of molecules, which is a basic process of statistical physics, are governed by electromagnetic forces. Then, it is natural to accept that the theory of heat and thermal processes must {\em relativistic}, i.e. formulated in a four-dimensional space-time. Nevertheless, an application of the fundamental physical-mathematical concept of space-time to the theory of thermal phenomena was never done. Attempts to modify existed thermodynamic equations for 'relativistic' conditions served little progress to thermal physics. To make a mathematically consistent theory, which is universal in describing experimental data, these old equations must be replaced by new equations, with new variables, based on new mathematics.

From the classic textbooks \cite{Landau-V5} to the modern tutorial \cite{Skacej-Ziherl-book2019}, the union of statistical physics and thermodynamics remains a physical theory that rests on the one hundred year old foundation and aims at solving one hundred year old problems. Contemporary experimental physics challenges orthodox theories, thus, theoretical physicists must respond by developing theories that would match modern mathematics with new and old experimental data. Presented above is the outline of a direction that may become, after critical consideration and further development, a working theory for thermal phenomena.

The modern system of scientific research, which over the last several decades became truly corporate-like, designed to produce a 'scientist-bureaucrat' \cite{Velazquez-book2019}. However, no true scientist can be a bureaucrat, and this is likely the reason why some obvious and urgent problems of physics have not be attacked for so long. The discrepancy between experiment and theory became very deep, and it requires an immediate action by the scientific community to close the gap and gain fast scientific progress.

{\em On new statistical physics.}
Even though old mathematical methods failed to complete the theory of thermal phenomena, the one one, based on the century old result in the number theory \cite{Hardy-PLMS1918}, can do it \cite{Maslov-MN2016,Maslov-MN2016-2,Maslov-RJMP2008,Maslov-MN2013}. The logical mistake done in the derivation of modern statistical physics \cite{Landau-V5} was uncovered by V.P. Maslov long time ago \cite{}. The correct combinatorics of indistinguishable objects in molecular physics allowed him to derive a new statistical distribution that can successfully describe properties of real gases. This is a natural development of science: new mathematics and new physics develop together as a unique physical-mathematical science, as envisioned and implemented by I. Newton, H. Poincare, A. Kolmogorov and many other great mathematical physicists. In this specific case, there was an unacceptably long delay with the application of new mathematics. It is  sad that this delay was at least partially caused by psychological and social reasons, like some personal attitudes towards the probability theory and the number theory.

{\em Testing the theory with experimental data}.
Let us quote J.W. Gibbs: {\em ''It is well known that while theory would assign to the gas six degrees of freedom per molecule, in our experiments on specific heat we cannot account for more than five. Certainly, one is building on an insecure foundation, who rests his work on hypotheses concerning the constitution of matter''}, \cite{Gibbs-book1902}, pp. ix–xi. Since that time the discrepancy between theory and experiment grew much larger however, an {\em apparent} agreement was reached by introducing new parameters and making a physical theory more complex, e.g. \cite{Vukalovich-book1953}. Clearly such a theory can be only regarded as a model, for it is not universal and cannot describe even one class of gases by the same equation
In the following paper we take as an example the simplest case of one-atomic gases, and noble gases are excellent representatives of ideal mono-atomic gases.



\end{document}